\documentclass[12pt]{article}
\pdfoutput=1
\usepackage{jheppub}
\usepackage{verbatim}
\usepackage{graphics, color}
\usepackage{graphicx}
\usepackage{amsmath}
\usepackage{amssymb}
\usepackage{amsthm}
\usepackage{amsfonts}
\usepackage[usenames,dvipsnames]{xcolor}
\usepackage[normalem]{ulem}

\newcommand{\be}{\begin{equation}}
\newcommand{\ee}{\end{equation}}

\newcommand{\tr}{\mathrm{tr}}

\newcommand{\HC}{\mathcal{H}_{\mathcal{C}}}
\newcommand{\lan}{\langle}
\newcommand{\ran}{\rangle}

\newcommand{\mO}{\mathcal{O}}

\newcommand{\wt}{\widetilde}

\newcommand{\W}{\mathcal{W}}
\newcommand{\J}{\mathcal{J}}
\newcommand{\ol}{\overline}

\newtheorem*{theorem}{Theorem}

\definecolor{grey}{rgb}{.5,.5,.5}
\definecolor{bluegreen}{rgb}{0,.5,.5}
\definecolor{darkgreen}{rgb}{0,.5,0}

\begin{document}

\title{Bulk Locality and Quantum Error Correction in AdS/CFT}
\author[a]{Ahmed Almheiri,} 
\author[a]{Xi Dong,}
\author[b]{Daniel Harlow}
\affiliation[a]{Stanford Institute for Theoretical Physics, Department of Physics, Stanford University, Stanford, CA 94305, USA}
\affiliation[b]{Princeton Center for Theoretical Science, Princeton University, Princeton NJ 08540 USA}
\emailAdd{almheiri@stanford.edu}
\emailAdd{xidong@stanford.edu}
\emailAdd{dharlow@princeton.edu}
\abstract{We point out a connection between the emergence of bulk locality in AdS/CFT and the theory of quantum error correction.  Bulk notions such as Bogoliubov transformations, location in the radial direction, and the holographic entropy bound all have natural CFT interpretations in the language of quantum error correction.
We also show that the question of whether bulk operator reconstruction works only in the causal wedge or all the way to the extremal surface is related to the question of whether or not the quantum error correcting code realized by AdS/CFT is also a ``quantum secret sharing scheme'', and suggest a tensor network calculation that may settle the issue.  Interestingly, the version of quantum error correction which is best suited to our analysis is the somewhat nonstandard ``operator algebra quantum error correction'' of Beny, Kempf, and Kribs. Our proposal gives a precise formulation of the idea of ``subregion-subregion'' duality in AdS/CFT, and clarifies the limits of its validity.}
\maketitle
\section{Introduction}
Almost twenty years after its initial formulation, the AdS/CFT correspondence remains our best-understood example of a precise theory of quantum gravity.  It has shed light on many deep puzzles in quantum gravity, and has also been of practical use in studying the dynamics of strongly interacting quantum field theories.  One aspect that remains mysterious, however, is the emergence of approximate bulk locality.  Locality near the boundary is straightforward. In the ``extrapolate'' version of the AdS/CFT dictionary we have a simple relation \cite{Banks:1998dd,Harlow:2011ke}
\be\label{extd}
\lim_{r\to \infty}r^{\Delta}\phi(r,x)=\mO(x)
\ee
between limiting values of a bulk field $\phi$ and a conformal field theory operator $\mO$; this dictionary manifestly respects locality in the $x$ directions since the CFT does.  The radial direction, however, is more subtle. One way to see this is to observe that naively a local operator in the center of the bulk should commute with every local operator at the boundary on a fixed time slice containing that bulk operator. This is not consistent, however, with a standard property of quantum field theory; any operator that commutes with all local operators at a fixed time must be proportional to the identity.\footnote{In lattice theories with scalars and fermions coupled to abelian gauge fields this property is essentially obvious in the Hamiltonian formulation.  Showing it for non-abelian gauge fields on the lattice requires more work.  In either case the idea is to show that the algebra generated by local operators on a time-slice acts irreducibly on the Hilbert space; the statement then follows from Schur's lemma.  In the continuum this idea is called the ``time-slice axiom'' \cite{Streater:1989vi,Haag:1992hx}; for recent rigorous discussions, see, e.g., \cite{Chilian:2008ye,Benini:2013ita}.  There are actually topological theories where the time-slice axiom is false, for example in Chern-Simons theory quantized on a topologically nontrivial Riemann surface, but we don't expect this loophole to be relevant for CFTs with ordinary gravity duals.
}
Bulk locality thus cannot be respected within the CFT at the level of the algebra of operators;  we'd then like to know in what sense it \textit{is} respected.\footnote{One subtlety in this argument is that to put the operator at a definite bulk point in a diffeomorphism-invariant way we need to include ``gravitational dressing'' that will allow the operator to not necessarily commute with local operators at the boundary at subleading order in $1/N$.  We will discuss this more in section \ref{backreactsec}, where we will see that this level of non-locality is not enough to avoid a contradiction between the bulk and boundary algebras.}  

The basic idea of this paper is that bulk locality is a statement about certain subspaces of states in the CFT.  That these subspaces can be large is a consequence of the large-$N$ properties of the CFT, but the large degree of non-local entanglement in finite energy states of the CFT also plays an essential role.  Our strategy will be to gradually back away from the $r\to \infty$ limit in equation \eqref{extd} and study how the the CFT representations of bulk operators spread in spatial support as we do so.  On the bulk side the tool we will mostly use is the AdS-Rindler reconstruction of bulk fields introduced in \cite{Hamilton:2006az} and refined in \cite{Morrison:2014jha}.  We will observe that this construction has several paradoxical features, which we will illuminate by recasting it on the CFT side in the language of quantum error correcting codes \cite{shor1995scheme,Gottesman:1997zz}.  This language gives a new, more general perspective on the issue of bulk reconstruction, and we believe that it is the natural framework for understanding the idea of ``subregion-subregion'' duality \cite{Bousso:2012sj,Czech:2012bh,Bousso:2012mh,Hubeny:2012wa}.  In particular, the radial direction in the bulk is realized in the CFT as a measure of how well CFT representations of bulk quantum information are protected from local erasures.  The holographic principle also naturally arises in the guise of the general statement that there is an upper bound on how much quantum information a given code can protect from erasures.

One point that will appear in this analysis is that truncated subalgebras of bulk observables are of interest; these were also advocated in \cite{Papadodimas:2013jku} in the context of describing the black hole interior.  Aspects of our proposal are inspired by their construction, but here we do not discuss black hole interiors and we are not violating quantum mechanics \cite{Harlow:2014yoa}.  A connection between black holes and quantum error correction was also made in \cite{Verlinde:2012cy}, which is essentially an earlier version of the proposal of \cite{Papadodimas:2013jku}, but again the context was different and our work here should be uncontroversial by comparison.

Before proceeding, let us establish a few conventions used throughout this paper.  We will frequently discuss subspaces and tensor factors of the Hilbert space.  When we say that an operator acts within a subspace we also mean that the same is true for its hermitian conjugate.  If the Hilbert space is a tensor product $\mathcal{H}_E \otimes \mathcal{H}_{\ol{E}}$, for any operator $O_{\ol{E}}$ that acts on $\ol{E}$ we may trivially form an operator $I_E \otimes O_{\ol{E}}$ that acts on the entire Hilbert space.  We will often drop the identity operator $I_E$ and write it simply as $O_{\ol{E}}$.  The reader should also keep in mind that the CFT regions $A$ and $\ol{A}$ that we will talk about in section \ref{adssec} unfortunately correspond to $\ol{E}$ and $E$ respectively in the language of section \ref{qcorrsec}.

\section{Bulk reconstruction and an AdS-Rindler puzzle}
\subsection{Global AdS reconstruction}\label{globalsec}
\begin{figure}
\begin{center}
\includegraphics[height=7cm]{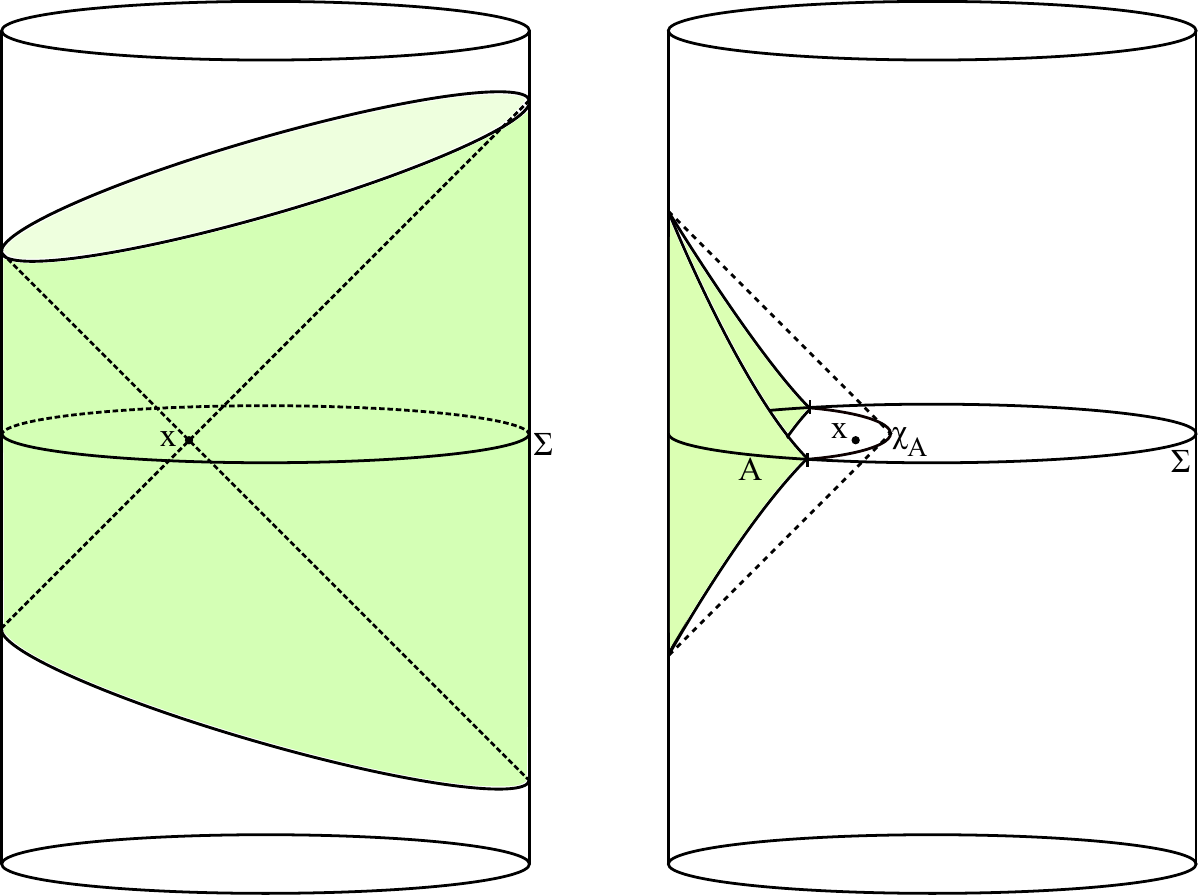}
\caption{$AdS_3$ reconstruction globally, and in an AdS-Rindler wedge.}\label{globalfig}
\end{center}
\end{figure}

We begin by briefly recalling the standard CFT construction of local bulk fields in AdS \cite{Banks:1998dd,Hamilton:2006az,Heemskerk:2012mn}.  We will first work in global coordinates, where the metric asymptotically has the form
\be\label{adsmetr}
ds^2\sim-(r^2+1)dt^2+\frac{dr^2}{r^2+1}+r^2 d\Omega_{d-1}^2.
\ee
The CFT dual to this system lives on $\mathbb{S}^{d-1}\times \mathbb{R}$, with the $\mathbb{R}$ being the time direction.  The Hilbert space of states is the set of field configurations on $\mathbb{S}^{d-1}$.  The idea is then to perturbatively construct operators in the CFT which obey the \textit{bulk} equations of motion, with the boundary conditions set by the dictionary \eqref{extd}.  For simplicity we will assume that all bulk interactions are suppressed by inverse powers of a quantity $N$, which will also set the AdS radius in Planck units. At leading order in $1/N$, this procedure results in a straightforward prescription for the CFT representation of a bulk field $\phi(x)$; we simply have
\be\label{smear}
\phi(x)=\int_{\mathbb{S}^{d-1}\times \mathbb{R}} dY K(x;Y)\mathcal{O}(Y),
\ee
where the integral is over the conformal boundary and $K(x,Y)$ is a so-called ``smearing function''.  The smearing function obeys the bulk wave equation in its $x$ index, and leads to \eqref{extd} as we take $x$ to the boundary.  It can be chosen to only have support when $x$ and $Y$ are spacelike separated, which we illustrate for $AdS_3$ in the left diagram of figure \ref{globalfig}; the point $x$ is represented by a boundary integral over the green region only.  In the case of empty AdS, where we take \eqref{adsmetr} to hold everywhere, explicit representations of the smearing function can be found in \cite{Hamilton:2006az,Heemskerk:2012mn}.\footnote{One subtlety here is that for more general asymptotically AdS backgrounds, we are not aware of a rigorous argument for the existence of $K$, even in the distributional sense that we will see we need to allow in the following subsection.  One obvious problem is that $x$ could be behind a horizon, but even for geometries with no horizons the only precise argument for the existence of $K$ (or more precisely the existence of the ``spacelike Green's function'' it is built from) requires spherical symmetry \cite{Heemskerk:2012mn}.  We are not aware of any obstruction to its existence, but it would nonetheless be very interesting to see a detailed analysis of this somewhat nonstandard problem in partial differential equations.}  $1/N$ corrections can be systematically included \cite{Kabat:2011rz,Heemskerk:2012mn}, although we won't really need to discuss them here.  At higher orders in this perturbation theory we will need to confront the problem of defining local operators in a diffeomorphism invariant theory, but we postpone discussion of this until section \ref{backreactsec}.  

It is not obvious from the definition that the operators \eqref{smear} have the expected commutators in the bulk; this has been checked perturbatively within low point correlation functions in \cite{Kabat:2011rz}, but must eventually break down in states with enough excitations to avoid a contradiction with the argument in our introduction.  We will argue below that, within the subspace of states that are ``perturbatively close'' to the vacuum, it breaks down only at the level of non-perturbatively small corrections.

Note that once we have a representation of the form \eqref{smear}, we can use the CFT Hamiltonian to re-express all operators on the right hand side in terms of Heisenberg picture fields on a single Cauchy surface in the CFT, denoted as $\Sigma$ in figure \ref{globalfig}.  This representation is quite nontrivial, in general it involves severely nonlocal and multitrace operators.  It also has the property that if we take $x$ to be near the boundary but not quite on it, the single-time CFT representation of $\phi(x)$ still involves operators with support on all of $\Sigma$.  We might hope to find a representation whose boundary support shrinks as the operator approaches the boundary, and indeed the AdS-Rindler representation does exactly this, as we will now explain.  

\subsection{AdS-Rindler reconstruction}\label{rindsec}
Consider a subregion $A$ of a CFT Cauchy surface $\Sigma$.  The \textit{boundary domain of dependence of $A$}, denoted $D[A]$, is defined as the set of points on the boundary with the property that every inextendible causal curve, meaning a curve whose tangent vector is never spacelike and which is not part of a larger curve with this property, that passes through it must also intersect $A$.  This is illustrated for the boundary of $AdS_3$ in the right diagram of figure \ref{globalfig}, where $A$ is the boundary interval lying between the two vertical hash marks and $D[A]$ is shaded green.  For any boundary region $R$, its \textit{bulk causal future/past} $\J^{\pm}[R]$ is defined as the set of bulk points which can be reached by bulk causal curves evolving from/to the region $R$.  The \textit{causal wedge of a CFT subregion A} \cite{Hubeny:2012wa} (for earlier related definitions see \cite{Bousso:2001cf}) is defined as
\be
\W_C[A]\equiv \J^+[D[A]]\cap\J^{-}[D[A]].
\ee
In the right diagram of figure \ref{globalfig}, $\W_C[A]$ roughly lies between the dashed lines and $D[A]$. The bulk codimension-two surface $\chi_A$ in the figure is the ``rim'' of the wedge and is commonly referred to as the \emph{causal surface} of $A$ \cite{Hubeny:2012wa}; more precisely it is defined as the part of the intersection of the boundaries of $\J^\pm[D[A]]$ that does not also intersect the conformal boundary at infinity.  $\chi_A$ can also be described as the intersection of the past and future horizons of $D[A]$.

A simple example of these definitions is where we take the geometry to be pure $AdS_{d+1}$, $\Sigma=\mathbb{S}^{d-1}$ to be the $t=0$ slice of the boundary, and $A$ to be one hemisphere of $\Sigma$.  In this case $\W_C[A]$ becomes what is usually referred to as the AdS-Rindler wedge.   
\begin{figure}
\begin{center}
\includegraphics[height=5cm]{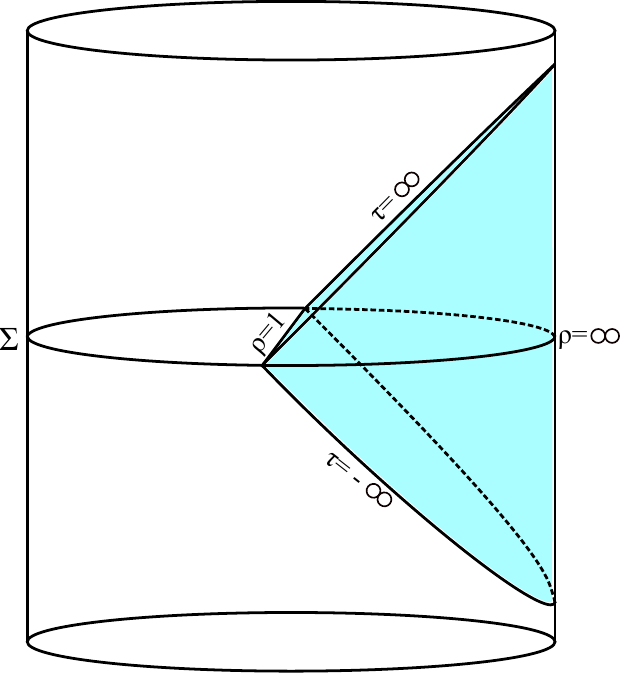}
\caption{Coordinates for the AdS-Rindler wedge for $AdS_3$, shaded in blue.  In this case we have $-\infty<x<\infty$.}\label{rindlerfig}
\end{center}
\end{figure}
A natural set of bulk coordinates on the AdS-Rindler wedge gives a metric with the form
\be
ds^2=-(\rho^2-1)d\tau^2+\frac{d\rho^2}{\rho^2-1}+\rho^2\left(dx^2+\sinh^2 x d\Omega_{d-2}^2\right),
\ee
where the coordinate ranges are $\rho>1, x\geq 0,-\infty<\tau<\infty$ and the geometry in parentheses is just the $d-1$ dimensional hyperbolic disc.  The causal surface $\chi_A$ is given by the limit $\rho \to 1$ at fixed $\tau$, and $A$ itself is given by $\rho\to \infty$ and $\tau=0$.  We illustrate this for $AdS_3$ in figure \ref{rindlerfig}.  By acting on this example with bulk isometries (or equivalently boundary conformal transformations), we can arrive at the causal wedge for any round disc in $\Sigma$.  The case of $AdS_3$ is especially simple; all connected boundary regions are intervals and thus can be produced in this way.  

The point then is that the construction of CFT representations of bulk fields in the previous subsection can also be implemented purely within the causal wedge \cite{Hamilton:2006az}.\footnote{This has been worked out explicitly only at leading order in $1/N$ for the case of the AdS-Rindler wedge, the $1/N$ corrections should basically be treatable using the same methods as for the global construction and the existence for more general geometries has the same caveats as before.} At leading order in $1/N$, the claim is that for any $\phi(x)$ with $x\in \W_C[A]$, we can again represent $\phi(x)$ via the expression \eqref{smear}, but with the $Y$ integral now taken only over $D[A]$.  This is illustrated for $AdS_3$ in the right diagram in figure \ref{globalfig}, where we have allowed for a conformal transformation that changes the size of the boundary interval $A$.  We review more details of this construction in appendix \ref{rindapp}; the only major subtlety is that the smearing function $K$ no longer exists as a function and must be understood as a distribution for integration against CFT expectation values \cite{Morrison:2014jha} (see also \cite{Papadodimas:2012aq} for some related discussion).  

Thus we see that the AdS-Rindler construction of $\phi$ indeed has the property that if $x$ is close to the boundary, only a small boundary region $A$ localized near $x$ is needed to be able to reconstruct $\phi$ in $D(A)$.  Moreover, by making use of the CFT evolution we can again rewrite the expression \eqref{smear} entirely in terms of nonlocal Heisenberg operators acting at $t=0$, but now they will act only on $A$.  

\subsection{Overlapping wedges}
\begin{figure}
\begin{center}
\includegraphics[height=4cm]{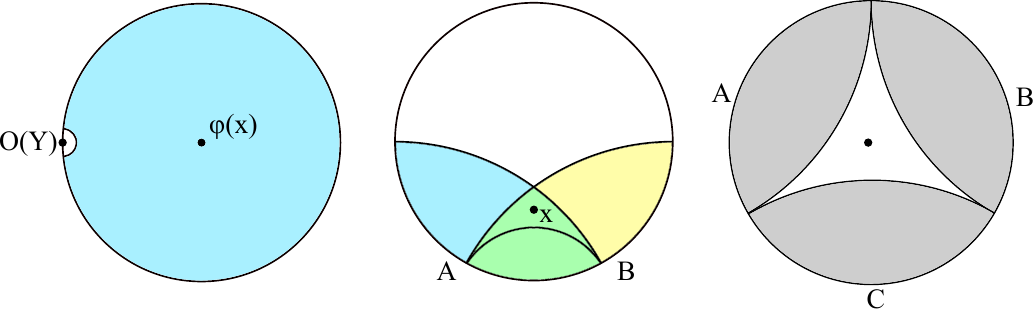}
\caption{Three examples of $AdS_3$-Rindler reconstruction.  Shown here is a top-down view of a bulk Cauchy slice whose boundary is $\Sigma$.  On the left, the blue shaded region is the intersection of this Cauchy slice with the causal wedge for a CFT region $A$ that is the complement of a small boundary interval around the boundary point $Y$.  In the center we have the point $x$ lying in the causal wedge of two different CFT regions, $A$ and $B$.  $A$ borders the blue and green regions, while $B$ borders the green and yellow regions. The black circle segments are $\chi_A$, $\chi_B$, and $\chi_{A\cap B}$.  On the right we have split $\Sigma$ into a union of three disjoint intervals, $A$, $B$, and $C$, and the circle segments are $\chi_A$, $\chi_B$, and $\chi_C$.}\label{paradox1fig}
\end{center}
\end{figure}
The AdS-Rindler construction of bulk fields we have just described has the somewhat counter-intuitive property that the same bulk field operator $\phi(x)$ lies in multiple causal wedges, and thus can be represented as an operator on distinct regions $A,B,\ldots$ in $\Sigma$.  One consequence of this is shown in the left diagram of figure \ref{paradox1fig}; for any bulk field operator $\phi(x)$ and any CFT local operator $\mO(Y)$ such that $x$ and $Y$ are spacelike separated, we can choose a causal wedge $\W_C[A]$ such that $\mO(Y)$ lies in the complement of $A$ in $\Sigma$.  By CFT locality $\mO(Y)$ then must exactly commute with our representation of $\phi(x)$ in that wedge.  This is coming dangerously close to contradicting the theorem mentioned in the introduction, that is that no nontrivial operator in the CFT can commute with all local CFT operators on $\Sigma$.  

To avoid this contradiction it must be the case that the representations of $\phi(x)$ in different wedges are not really all the same operator on the CFT Hilbert space.  We can see this in another way by considering the setup of the center diagram in figure \ref{paradox1fig}, where we have two overlapping wedges $\W_C[A]$ and $\W_C[B]$ that both contain the point $x$ but $x$ is not contained in $W_C[A\cap B]$.  For a CFT operator defined with support only on $A$ to really be equal to a CFT operator defined with support only on $B$, it must be that the operator really only has support on $A\cap B$.  But given that we have chosen $x$ to lie outside of $W_C[A\cap B]$, we do not expect the operator to have such a representation.  In fact in this example the operator has a representation on the \textit{complement} of $A \cap B$, and we will see in section \eqref{opalgsec} that when this is so a version of the no-cloning theorem of quantum mechanics forbids an accurate representation of the operator on $A\cap B$.

We can see the non-equivalence of the operators even more clearly by considering a third example, shown in the right diagram in figure \ref{paradox1fig}.  Now a bulk field at the point $x$ lies outside of the causal wedge for any one of the regions, but it can be reconstructed in $A \cup B$, $B \cup C$, or $A \cup C$.  The mutual intersection of these regions is just three points, and if we consider another set of three regions slightly rotated from these we can come up with a set of six possible reconstructions whose mutual intersection is genuinely empty.  There is simply no possible way that they can all be equal as operators.  For future reference we will refer to the three operators as $\phi_{AB}(x)$, $\phi_{BC}(x)$, and $\phi_{AC}(x)$.  

We thus need to decide how we are to reconcile these operator inequivalences with the fact that in the bulk theory it seems that the operators are equivalent.  There will clearly be some CFT states where they act quite differently, and we would like to understand the physics of the subset of states where their action is equivalent.  This problem can be nicely understood in the language of quantum error correction, to which we now turn.

\section{Correcting quantum erasures}\label{qcorrsec}
Say Alice wants to send Bob a quantum state of $k$ qubits in the mail, but she is worried that some of the qubits might get lost on the way.  Quantum error correction is a procedure that allows her to embed this state into $n>k$ qubits in such a way that even if some qubits are lost, Bob can still recover it.  In this section we review some basic facts about this, beginning with an example.\footnote{Our presentation of quantum error correction is somewhat nonstandard, since we are interested only correcting for the erasure of a known set of qubits. This allows us to omit many of the usual topics, such as quantum channels, ancilla, check operators, etc.  Standard reviews of this more general formalism are \cite{preskillnotes,nielsen2010quantum}; for a concise description of the basic ideas see section 4 of \cite{Harlow:2013tf}.}

\subsection{A simple example of erasure correction}\label{examplesec}
The simplest example of quantum error correction actually involves three-state ``qutrits'' instead of two-state qubits, and it uses three qutrits to send a single-qutrit message \cite{Cleve:1999qg}.  Say Alice wishes to send the state
\be
|\psi\ran=\sum_{i=0}^2 a_i |i\ran.  
\ee
The idea is to instead send the state
\be
|\wt{\psi}\ran=\sum_{i=0}^2 a_i |\wt{i}\ran,
\ee
where
\begin{align}\nonumber
|\wt{0}\ran&=\frac{1}{\sqrt{3}}\left(|000\ran+|111\ran+|222\ran\right)\\\label{codespace}
|\wt{1}\ran&=\frac{1}{\sqrt{3}}\left(|012\ran+|120\ran+|201\ran\right)\\\nonumber
|\wt{2}\ran&=\frac{1}{\sqrt{3}}\left(|021\ran+|102\ran+|210\ran\right).
\end{align} 
This protocol has two remarkable properties.  First of all for any state $|\wt{\psi}\ran$, the reduced density matrix on any one of the qutrits is maximally mixed.  Thus no single qutrit can be used to acquire any information about the state.  Secondly, from any two of the qutrits Bob \text{can} reconstruct the state.  For example, say he has access to only the first two qutrits.  He can make use of the fact that there exists a unitary transformation $U_{12}$ acting only on the first two qutrits that implements 
\be\label{U12}
\left(U_{12}\otimes I_3\right)|\wt{i}\ran=|i\ran\otimes \frac{1}{\sqrt{3}}\left(|00\ran+|11\ran+|22\ran\right).
\ee
Acting with this on the encoded message, we see that Bob can recover the state $|\psi\ran$: 
\be
\left(U_{12}\otimes I_3\right)|\wt{\psi}\ran=|\psi\ran\otimes \frac{1}{\sqrt{3}}\left(|00\ran+|11\ran+|22\ran\right).
\ee
Explicitly $U_{12}$ is a permutation that acts as
\be
\begin{tabular}{ l c r }
$|00\ran\to|00\ran$ & $|11\ran \to |01\ran$ & $|22\ran\to |02\ran$ \\
$|01\ran \to |12\ran$ & $|12\ran \to |10\ran$ & $|20\ran \to |11\ran$ \\
$|02\ran\to |21\ran$ & $|10\ran\to|22\ran$ & $|21\ran \to |20\ran$ \\
\end{tabular}.
\ee
Clearly by the symmetry of \eqref{codespace} a similar construction is also possible if Bob has access only to the second and third, or first and third qutrits.  Thus Bob can correct for the loss of any one of the qutrits; in quantum information terminology one describes this as a quantum error correcting code that can protect against arbitrary single qutrit erasures.  The subspace spanned by \eqref{codespace} is called the \textit{code subspace}; the entanglement of the states in the code subspace is essential for the functioning of the protocol.

In our discussion of reconstruction in the previous section we were interested in the action of operators rather than the recovery of states, and we can rephrase the error correction protocol in this language.  Indeed, say that $O$ is an operator that acts on the single qutrit Hilbert space as\footnote{Here we write $(O)_{ij}$ to indicate the matrix elements of the operator $O$ on the code subspace, with the parentheses there to distinguish this from the operators $O_{12}$, $O_{23}$, etc to be defined momentarily.} 
\be\label{Odef}
O|i\ran=\sum_{j}(O)_{ji}|j\ran.
\ee
For any such $O$ we can always find a (non-unique) three-qutrit operator $\wt{O}$ which implements the same transformation on the code subspace:
\be
\wt{O}|\wt{i}\ran=\sum_{j}(O)_{ji}|\wt{j}\ran.
\ee
In quantum computing language, operators like $\wt{O}$ that act directly on the code subspace in this manner are called \textit{logical operations}, since they are the types of things that we want to implement when performing a fault-tolerant quantum computation. 

For a general code subspace, $\wt{O}$ would need to have nontrivial support on all three qutrits.  For the code subspace in question, however, is straightforward to see that the operator
\be
O_{12}\equiv U_{12}^\dagger O U_{12},
\ee
where $O$ is taken to act on the first qutrit, acts as
\be
O_{12}|\wt{i}\ran=\sum_j (O)_{ji}|\wt{j}\ran.
\ee
$O_{12}$ is thus an $\wt{O}$ that has support only on the first two qutrits.  Since we can also analogously construct $O_{23}$ or $O_{13}$, we have realized a situation where operators with nontrivial support on different qutrits have the same action on the code subspace.  This should be reminiscent of our discussion of overlapping wedges in the previous section; we will make the connection more explicit soon but first we need to discuss some general properties of quantum erasure correction.

Before moving on, however, we want to introduce a notational simplification.  So far we have been careful to distinguish the single-qutrit operator $O$ from its three-qutrit representations $\wt{O}$.  We find it convenient, however, to from now on abuse notation by instead thinking of ``$O$'' as an abstract logical operation and using it both cases; which operator we mean should always be clear from the context.  So for example we can write
\be
O_{12}|\wt{i}\ran=O|\wt{i}\ran.
\ee

\subsection{General erasure correction}\label{errcorsec}
We now describe a natural generalization of the protocol of the previous subsection. For familiarity we will describe it using qubits, although none of the results rely on this.  Say that we want to protect a $k$-qubit code subspace of an $n$-qubit system against the loss of some collection of $E$ of $l$ of the qubits.  We define the code subspace $\HC$ as the span of the orthonormal states
\be\label{codesubspace}
|\wt{i}\ran=U_{enc}|i_1 \ldots i_k 0_{k+1}\ldots 0_{n}\ran,
\ee
where $U_{enc}$ is called the encoding unitary transformation.  There is a necessary and sufficient condition for the correctability of the erasure of $E$ \cite{PhysRevA.54.2629}.  Say that we adjoin to our system a reference system $R$ of $k$ additional qubits.  We then consider the state
\be
|\phi\ran=2^{-k/2}\sum_i |i\ran_R |\wt{i}\ran_{\overline{E}E},
\ee
where $\overline{E}$ denotes the set of $n-l$ qubits that aren't erased.  The code \eqref{codesubspace} can correct for the erasure of $E$ if and only if we have
\be\label{erasecond}
\rho_{RE}[\phi]=\rho_{R}[\phi]\otimes \rho_{E}[\phi].
\ee
Here $\rho_R[\phi]$, $\rho_E[\phi]$, etc are the reduced density matrices obtained from $|\phi\ran$ by partial trace.  This is equivalent to saying that the \textit{mutual information} $I_{RE}=S_R+S_E-S_{RE}$ vanishes, where $S_X$ is the Von Neumman entropy, $S_X\equiv-\tr \rho_X \log \rho_X$.  Let us first see that this ensures we can correct the erasure.  The Schmidt decomposition\footnote{The Schmidt decomposition of a pure state $|\psi\ran$ in a bipartite Hilbert space $\mathcal{H}_A\otimes \mathcal{H}_B$ is the observation that for any $|\psi\ran\in \mathcal{H}_A\otimes \mathcal{H}_B$ there exists a set of orthonormal states $|i\ran_A$ in $\mathcal{H}_A$, a set of orthonormal states $|i\ran_B$ in $\mathcal{H}_B$, and a set of non-negative real numbers $C_i$ such that
\be
|\psi\ran=\sum_i C_i |i\ran_A |i\ran_B.
\ee
For a derivation and some more details see for example \cite{Harlow:2014yka}.} of $|\phi\ran$, together with \eqref{erasecond}, ensures us that there exists a basis $|e\ran$ for $E$ and a set of orthonormal states $|\psi_{i,e}\ran_{\overline{E}}$ in $\overline{E}$ such that
\be
|\phi\ran=2^{-k/2}\sum_{i,e}C_e|i\ran_R |e\ran_E |\psi_{i,e}\ran_{\overline{E}}, 
\ee
where $C_e$ are some non-negative coefficients obeying $\sum_e C_e^2=1$.  In other words there exists a unitary transformation $U_{\overline{E}}$ acting only on $\overline{E}$ such that
\be
U_{\overline{E}}|\phi\ran=2^{-k/2}\sum_i |i\ran_R|i\ran_{\overline{E}_1}\otimes |\chi\ran_{\overline{E}_2 E},
\ee
where we have denoted the first $k$ qubits of $\overline{E}$ as $\overline{E}_1$ and the rest as $\overline{E}_2$.  $|\chi\ran$ is some state that is independent of $i$.  This then implies that we must have
\be
U_{\overline{E}}|\wt{i}\ran_{\overline{E}E}=|i\ran_{\overline{E}_1}\otimes |\chi\ran_{\overline{E}_2 E},
\ee
which is the analogue of \eqref{U12} above and demonstrates that we can use $U_{\overline{E}}$ to correct the erasure.  If we do not have \eqref{erasecond}, then there is nonzero correlation between $R$ and $E$, so we can learn about the state of $R$ by doing measurements on $E$.  Since any successful protocol must not care about what happens to the qubits we lose, this prevents us from being able to correct the erasure.  We can thus loosely rephrase \eqref{erasecond} as the statement that the erasure of $E$ is correctable if and only if no information about $i$ can be obtained from $E$.  This is related to the no-cloning theorem; if we were able to get the same quantum information about the encoded state $|\wt{\psi}\ran$ from both $E$ and $\overline{E}$ then we would have built a machine for cloning that information.

There is a useful reformulation of the condition \eqref{erasecond} as the statement that for any operator $X_E$ acting on $E$, we must have \cite{grassl1997codes}
\be\label{erasecond2}
\lan \wt{i}|X_E|\wt{j}\ran=\delta_{ij} C(X).
\ee
In other words we must have the projection of $X_E$ onto the code subspace be proportional to the identity.  One immediate consequence of this is that in any state $|\wt{\psi}\ran$ in the code subspace,  the correlation function of $X_E$ with any operator $O$ that acts within the code subspace\footnote{Throughout this paper, when we say that an operator acts within a subspace we mean that the same is true for its hermitian conjugate as well.} must vanish:
\be\label{codecorr}
\lan \wt{\psi}|O X_E|\wt{\psi}\ran-\lan \wt{\psi}|O|\wt{\psi}\ran\lan \wt{\psi}|X_E|\wt{\psi}\ran=0.
\ee
This is another manifestation of the idea that $E$ has no access to the encoded information.  

As in the previous subsection, we can use $U_{\overline{E}}$ to realize any operator $O$ acting within the code subspace as an operator $O_{\overline{E}}$ that acts just on $\overline{E}$.  Indeed we have both
\begin{align}\nonumber
O_{\overline{E}}|\wt{\psi}\ran&=O|\wt{\psi}\ran\\
O_{\overline{E}}^\dagger|\wt{\psi}\ran&=O^\dagger|\wt{\psi}\ran.\label{erasecond3}
\end{align}
In fact the converse of this statement also holds; if any operator on the code subspace can be realized as an operator on $\overline{E}$ as in \eqref{erasecond3}, then the code is able to correct for the loss of $E$.  The proof is simple. Say that the code were not correctable; then as just discussed there must exist an operator $X_E$ on $E$ where \eqref{erasecond2} does not hold.  By Schur's lemma, there must then exist an operator $O$ on the code subspace that does not commute with $X_E$ on $\HC$, that is with $\lan\wt{i}|[O,X_E]|\wt{j}\ran\neq 0$ for some $i$ and $j$.  But this operator $O$ can't be realized on $\HC$ by an operator $O_{\overline{E}}$ that acts only on $\overline{E}$, since any such operator by definition would commute with $X_E$.  
 
We now turn to the question of when we should expect \eqref{erasecond} (or equivalently \eqref{erasecond2} or \eqref{erasecond3}) to hold.  In situations where we would like our code to be able to correct against a wide variety of erasures, we expect that $\rho_E[\phi]$ will have full rank.\footnote{This excludes trivial cases like $U_{enc}=1$ with $E$ chosen to be the last $n-k$ qubits.  This code is completely defeated by erasing any of the first $k$ qubits, and if we knew that erasures would only affect the last $n-k$ qubits why would we include those qubits at all?}  In that case, in order to be able to have the orthonormal set of states $|\psi_{i,e}\ran$ we need the dimensionality of $\overline{E}$ to be at least as large as dimensionality of $RE$.  In other words we need
\be\label{lcond}
n\geq2l+k.
\ee
This condition is quite intuitive; wanting to send a larger message or correct larger erasures requires more qubits.  

In fact for large systems \eqref{lcond} is typically not only necessary but sufficient.  Say we take $|\phi\ran$ to be a random state of $2^{k+n}$ qubits in the Haar measure.  By Page's theorem \cite{Page:1993df}\footnote{Page's theorem is the statement that a random state of a bipartite Hilbert space $\mathcal{H}_A\otimes \mathcal{H}_B$ will be maximally mixed on the smaller factor up to corrections that go like the ratio of dimension of the smaller factor to the dimension of the larger factor.  For more details see e.g. \cite{Harlow:2014yka}.}, the density matrix of $R$ will be exponentially close to maximally mixed provided that $n-k\gg 1$, so by the Schmidt decomposition this is equivalent to choosing a random $k$-qubit code subspace of $n$ qubits.  The condition \eqref{erasecond} will hold if $RE$ is maximally mixed, which again by Page's theorem should be true provided that $n-2l-k \gg 1$. Thus, not only is \eqref{lcond} necessary for a typical code to correct for the loss of a particular set $E$ of $l$ qubits, it is basically sufficient for the code to correct for the loss of \textit{any} set of $l$ qubits.   

\subsection{Quantum secret sharing}\label{qsssec}
The three-qutrit example of section \ref{examplesec} has the interesting property that every collection of qutrits either can perfectly reconstruct the state $|\psi\ran$ or has no information about it at all.  General error correcting codes do not have this property, since sometimes we can have erasures which can be ``partially corrected'', but it is interesting to think about the codes that do. Say that we have a Hilbert space that is a tensor product of $p$ factors of not necessarily equal size, which in this context we will refer to as \textit{shares}.  A code subspace $\mathcal{C}$ of this Hilbert space is called a \textit{Quantum Secret Sharing Scheme} if it has the property that every collection of shares either can distinguish perfectly different elements of $\mathcal{C}$, meaning given access to it we can correct for the erasure of its complement, or it cannot distinguish different elements at all \cite{Cleve:1999qg}.  Collections which enable erasure correction are called \textit{authorized} and collections which do not are called \textit{unauthorized}.  We will see a possible application of quantum secret sharing to AdS/CFT in section \ref{discsec} below.

\subsection{Approximate erasure correction}\label{appxsec}
So far we have discussed exact quantum error correction, but in AdS/CFT we only expect the emergence of the bulk to be approximate.  It will thus be important for us to get a sense of how badly we might want to allow our three necessary and sufficient conditions for correctability to be violated.  The simplest way to relax the condition \eqref{erasecond} is to require only \cite{schumacher2002approximate}
\be\label{erasecondappx}
||\rho_{RE}-\rho_R\otimes \rho_E||_1\ll 1.
\ee
Here $||M||_1\equiv\mathrm{tr}\sqrt{M^\dagger M}$ is the trace norm of $M$; two density matrices whose difference has trace norm $\epsilon$ are ``operationally close'' in the sense that the probability distributions they predict for arbitrary measurements differ by at most $\epsilon$.  This essentially says that \textit{typical} states in the code subspace can be reconstructed to accuracy $\epsilon$; following \cite{schumacher2002approximate} we take this to be the definition of approximate error correction.  We would like to relate this to our second condition for correctability, \eqref{erasecond2}, but we need a convenient way to quantify the violation of \eqref{erasecond2}.  One good choice is to use correlation functions of the form
\be\label{phicorr}
C_\phi(O,X_E)\equiv\lan \phi|O^T_R X_E|\phi\ran-\lan\phi|O^T_R|\phi\ran\lan\phi|X_E|\phi\ran=\tr_{RE} \left[O^T_R X_E\left(\rho_{RE}-\rho_R\otimes\rho_E\right)\right].
\ee
Here $O^T_R$ denotes taking the transpose of an operator $O$ on the code subspace and acting with it on the reference system $R$; by construction acting on the state $|\phi\ran$ this is equivalent to acting with $O$ on $\ol{E}E$.  $C_\phi(O,X_E)$ is essentially the average of the correlation function \eqref{codecorr} over all $|\wt{\psi}\ran\in \HC$; they become equal in the limit of a large code subspace.\footnote{This average can be computed using the unitary integration technology described for example in appendix D of \cite{Harlow:2014yka}; the corrections vanish like powers of $2^{-k}$, which will be exponentially small in $N$ for our AdS-CFT construction.}  From the right hand side of \eqref{phicorr} it is not difficult to show that \cite{wolf2008area}
\be\label{cirac}
||\rho_{RE}-\rho_R\otimes \rho_E||_1\geq \Big|\frac{C_\phi(O,X_E)}{\lambda_O\lambda_X}\Big|,
\ee
where $\lambda_O$ and $\lambda_X$ are the largest eigenvalues of their respective operators.  Thus we see that, as one might expect from our discussion around \eqref{erasecond2}, the presence of nonzero correlation between $O$ and $X_E$ puts a limit on how accurately we can correct for the erasure of $E$.  This inequality will be very useful in our discussion of AdS/CFT, since after all computing correlation functions in the bulk theory is much easier than computing the trace norm directly.

\subsection{Operator algebra quantum error correction}\label{opalgsec}
In our discussion of AdS/CFT we will soon see that the presence of bulk correlation puts nontrivial restrictions on the correctability of errors via the inequality \eqref{cirac}.  There is, however, a generalized version of quantum error correction, called operator algebra quantum error correction, that is able to accommodate such correlation by requiring that our third necessary and sufficient condition \eqref{erasecond3} apply only to a \textit{subalgebra} of operators on the code subspace \cite{beny2007generalization,beny2007quantum}.  This requirement is greatly illuminated by the following theorem:
\begin{theorem}
Say that we have a code subspace $\HC \subset \mathcal{H}_E \otimes \mathcal{H}_{\ol E}$ and an operator $O$ that, together with its hermitian conjugate, acts within the code subspace.  In other words we have
\begin{align}\nonumber
O|\wt{i}\ran&=\sum_j O_{ji}|\wt{j}\ran \,,\\
O^\dagger|\wt{i}\ran&=\sum_j O^*_{ij}|\wt{j}\ran \,.
\end{align}
Then there exists an operator $O_{\ol{E}}$ acting just on $\ol{E}$ that obeys
\begin{align}\nonumber
O_{\ol{E}}|\wt{\psi}\ran&=O|\wt{\psi}\ran \,,\\
O_{\ol{E}}^\dagger|\wt{\psi}\ran&=O^\dagger|\wt{\psi}\ran \label{opalg}
\end{align}
for any $|\wt{\psi}\ran\in \HC$ if and only if $O$ commutes with the projection of any operator $X_E$ onto the code subspace, where $X_E$ acts on $E$.  In other words
\be\label{commreq}
\lan\wt{i}|[O,X_E]|\wt{j}\ran=0  \qquad \forall i,j.
\ee
\end{theorem}
We give a proof of this theorem in appendix \ref{thmapp}.  It is clear that the set of $O$'s that satisfy the assumptions of the theorem form a unital *-subalgebra $\mathcal{A}$ of the operators on the code subspace, meaning they include the identity and are closed under addition, multiplication, and hermitian conjugation.   If we take $\mathcal{A}$ to be the entire algebra of operators on $\HC$ then we recover our condition \eqref{erasecond3}.  Notice, however, that when $\mathcal{A}$ is a proper subalgebra we cannot use our previous argument to derive the condition \eqref{erasecond2} from \eqref{erasecond3}, since the $O$ that we constructed that doesn't have an $O_{\ol{E}}$ will not be in $\mathcal{A}$.  This gives a loophole that simultaneously allows correlation between $O$ and $X_E$ and the existence of $O_{\ol{E}}$.

For example in the two qubit system, consider a code subspace spanned by $\frac{1}{\sqrt{2}}\left(|00\ran+|11\ran\right)$ and $\frac{1}{\sqrt{2}}\left(|01\ran+|10\ran\right)$.  The operator $X$ that exchanges these two states can be realized just on the first qubit as the $X_1$ operator that flips it, even though in either state this operator is perfectly correlated with the $X_2$ operator that flips the second qubit.  This is possible because the $Z$ operator on the code subspace, for which the first state is a $+1$ eigenstate and the second state is a $-1$ eigenstate, cannot be realized as an operator just on the first qubit; this code corrects only the subalgebra generated by $1$ and $X$.  

This example has the perhaps surprising property that the encoded $1$ and $X$ operators can be realized on either of the two qubits, which seems in tension with our discussion of the no-cloning theorem above \eqref{erasecond2}.  This is an artifact, however, of the fact that this subalgebra is abelian, and is thus in some sense classical.  It is easy to prove that as long as the subalgebra is non-abelian, if it can be represented on $\ol{E}$ then it cannot be represented on $E$; the proof follows immediately by contradiction if we look at the commutator of two non-commuting elements of the algebra, but with one represented on $E$ and one represented on $\ol{E}$.  We used this ``algebraic no-cloning theorem'' above in our discussion of figure \ref{paradox1fig}.

\section{AdS/CFT as quantum error correction}\label{adssec}

\begin{figure}
\begin{center}
\includegraphics[height=5cm]{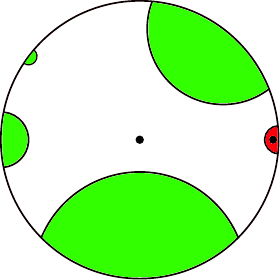}
\caption{Correcting for erasures in AdS/CFT.  Bulk quantum information at point in the center is protected in the CFT against the erasure of the boundary of any one of the green regions, but bulk information at the point near the boundary is completely lost by an erasure of the boundary of the red region.}\label{erasures}
\end{center}
\end{figure}

We now return to our discussion of bulk reconstruction.  Consider again the right diagram in figure \ref{paradox1fig}.  We argued using the AdS-Rindler reconstruction that the operator in the center can be represented either as an operator $\phi_{AB}$ with support on $A\cup B$, an operator $\phi_{BC}$ with support on $B\cup C$, or an operator $\phi_{AC}$ with support on $A \cup C$.  By now it should be obvious that this is directly analogous to the situation with $O_{12}$, $O_{23}$, and $O_{13}$ in the three qutrit example, or more generally the existence of the operator $O_{\ol{E}}$.  The main proposal of this paper is that this is more than an analogy, it is actually how AdS/CFT is reproducing the bulk!  In other words we can think of local bulk operators as logical operations on an encoded subspace, which becomes better and better protected against localized boundary errors as we move the operators inwards in the radial direction.\footnote{One detail here is that in defining the ``center'' of AdS we are implicitly choosing a conformal frame in the CFT; otherwise AdS is a homogeneous space and no point is special.  This is also implicit in our notion of ``small'' and ``large'' erasures in the CFT.}  We illustrate this in figure \ref{erasures}.  In the remainder of the paper we will spell out this idea in more detail, giving the bulk versions of most of the statements of the previous section.

\subsection{Defining code subspaces}
We begin by defining a set of candidate code subspaces for AdS/CFT. Our proposal is that we should pick some finite set of local bulk operators $\phi_i(x)$, realized in the CFT via the global representation of section \ref{globalsec}.  We then define a code subspace $\HC$ as the linear span of states of the form
\be\label{bulkcode}
|\Omega\ran, \phi_i(x)|\Omega\ran, \phi_i(x_1)\phi_j(x_2)|\Omega\ran,\ldots,
\ee
where we take the range of $i$, the number of $\phi_i(x)$'s we act with, and the number of points $x$ where the operators can be located to be bounded by some fixed finite number.  Here $|\Omega\ran$ is the ground state of the system; we could also do a similar construction around other sufficiently ``semiclassical'' states, but for rigor we will stick to $|\Omega\ran$ since, as mentioned in section \ref{globalsec}, the existence of appropriate smearing functions has not been completely established in the general case.  We postpone to section \ref{backreactsec} the question of how large $\HC$ can be.  It is essential that our definition of the code subspace will be different for different choices of the operators $\phi_i(x)$; the set of erasures that are correctable will depend on this choice, and we can learn about the way that the bulk theory is realized in the CFT by studying this dependence.  For example, in figure \ref{erasures} we see that moving the operators closer to the boundary makes our code subspace less protected against small erasures.  The CFT is not just one error-correcting code, it is many at once!  

We would like to think of the operators $\phi_i(x)$ as logical operations on this code subspace, but this does not quite work since by construction acting repeatedly with $\phi_i(x)$ will eventually take us out of $\HC$.  To get a set of operators that really act within $\HC$ we can include projection operators onto $\HC$ on both sides of $\phi_i$; these will be irrelevant except in studying high-point correlation functions, so we will not carry them around explicitly here.\footnote{A subtlety here is that the states \eqref{bulkcode} are not mutually orthogonal; a state of the form $\phi^m|\Omega\ran$ has a normalized overlap with the vacuum of order $\frac{\lan\Omega|\phi^m|\Omega\ran}{\sqrt{\lan\Omega|\phi^{2m}|\Omega\ran}}\sim 2^{-m}$.  The scaling with $m$ follows from counting Wick contractions at leading order in $1/N$.  If we choose our code subspace such that the maximum power of each $\phi$ grows like some power of $N$, then for states on which the projection onto $\HC$ is nontrivial this overlap will be exponentially small in $N$.  It then will be okay to conjugate the $\phi$'s by these projectors without ruining their low-point correlation functions.}  Now consider a decomposition of the boundary Cauchy surface $\Sigma$ into $A$ and $\ol{A}$.  If our code subspace $\HC$ can protect against the erasure of $\ol{A}$, then by our condition \eqref{erasecond3} it must be that we can find a representation of any operator on $\HC$ with support only in $A$.  In fact, this is what the AdS-Rindler reconstruction we reviewed in section \ref{rindsec} provides us; any causal wedge $\W_C[A]$ which contains the locations of the $\phi_i(x)$'s used in defining $\HC$ will allow a set of operators $\phi_{A,i}(x)$ with support only on $A$ and whose action on $\HC$ is the same as that of $\phi_{i}(x)$.  We now see that in the CFT this is a statement about being able to correct for the erasure of $\overline{A}$.

To avoid confusion, we stress that, just because we do not include some $\phi(x)$ in defining the code subspace, we do not mean to imply that its AdS-Rindler reconstruction does not work on that subspace.  We could easily consider a slightly larger subspace where we include it, and we could then interpret its AdS-Rindler reconstruction as arising from quantum error correction.  The only fundamental limitation on the AdS-Rindler reconstruction comes from the backreaction considerations we discuss in section \ref{backreactsec} below.

\subsection{Bulk correlation and smearing}\label{bcs}
It is illuminating to understand in more detail to what extent the AdS-Rindler reconstruction is consistent with our three equivalent conditions \eqref{erasecond}, \eqref{erasecond2}, \eqref{erasecond3} for quantum erasure correction.  We clearly do not expect them to hold exactly, but we might hope for them to hold in the approximate sense of \eqref{erasecondappx}.  As we explained in section \ref{appxsec}, a good diagnostic for approximate quantum erasure correction is that the correlation functions between operators acting within the code subspace and operators acting on the set to be erased are small enough that the inequality \eqref{cirac} does not preclude \eqref{erasecondappx} from holding.\footnote{We thank Juan Maldacena for pointing out the relevance of \eqref{cirac} in this situation.}  
\begin{figure}
\begin{center}
\includegraphics[height=5cm]{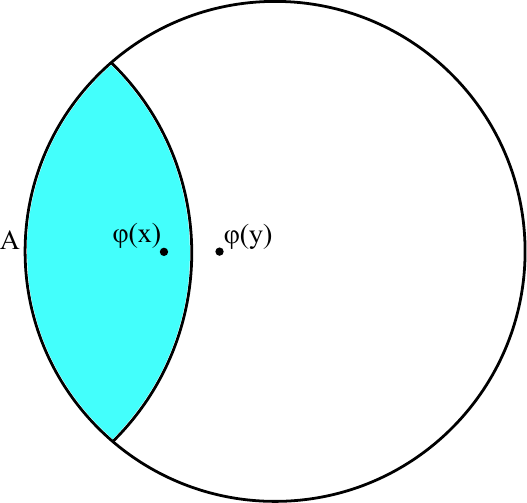}
\caption{Potentially troublesome bulk correlation.  Here $\phi(x)$ is an operator that acts within the code subspace $\HC$, and which we thus expect can be represented as an operator on $A$.  $\phi(y)$ we similarly expect to reconstructed on $\ol{A}$, but there is nonvanishing correlation between them in the ground state $|\Omega\ran$.  Using inequality \eqref{cirac}, this correlation puts a lower bound on the accuracy with which we can view AdS/CFT as quantum error correction in the conventional sense of section \ref{errcorsec}.}\label{corrfig}
\end{center}
\end{figure}

In fact it is a basic property of bulk physics that there is correlation between fields in $\W_C[A]$ and fields in $\W_C[\ol{A}]$, as we indicate in figure \ref{corrfig}. In deciding whether or not this bulk correlation interferes with our interpretation of AdS-Rindler reconstruction as quantum error correction, we need to properly take into account the operator eigenvalues in the denominator of \eqref{cirac}.  Formally these are infinite in a continuum quantum field theory, but every quantum field theorist knows that field operators are not really well-defined until they are integrated against smooth test functions with support over some region of nonzero measure, which we will take to have linear size $\epsilon_s$.  For simplicity we will take the bulk fields to be massless scalars and take their separation to be small compared to the AdS radius, in which case we have
\be\label{bulkcorr}
\frac{\lan \Omega|\phi(x)\phi(y)|\Omega\ran}{\lambda_\phi^2}\sim\left(\frac{\epsilon_s}{d(x,y)}\right)^{d-1}.
\ee
Here $d$ is the spacetime dimension of the boundary theory and $d(x,y)$ is the geodesic distance between $x$ and $y$.  This formula also holds in other states we produce by acting on $|\Omega\ran$ with smeared operators near $x$, and thus on average in the code subspace $\HC$.  We thus see that the right hand side of \eqref{cirac} will be small in our case provided that the operators $\phi_i(x)$ used in constructing $\HC$ are smeared over a distance which is small compared to their distance to the causal surface $\chi_A$ of the wedge $\W_C[A]$ in which we are trying to reconstruct them.  

This observation does much to justify our interpretation of AdS-Rindler reconstruction as quantum error correction, but it is somewhat unsatisfactory in the sense that the AdS-Rindler reconstruction still seems to work in the situation where we smear the operators over a distance that is comparable to their distance to the bifurcate Rindler horizon $\chi_A$, even though the bulk correlation is then too large to be ignored.  Indeed we interpret this as saying that the conventional quantum error correction of section \ref{errcorsec} does not fully capture the mechanism by which AdS/CFT realizes bulk locality.  The operator algebra quantum error correction introduced in section \ref{opalgsec}, however, provides precisely the generalization we need to fix this.  Consider for example an operator $S$ which acts on $\phi(x)|\Omega\ran$ as $S\phi(x)|\Omega\ran=|\Omega\ran$, and which annihilates any state orthogonal to $\phi(x)|\Omega\ran$.  This is an operator that acts within the code subspace, but its commutator with an operator $\phi(y)$ in $\W_C[\ol{A}]$ obeys
\be
\lan \Omega|[S,\phi(y)]|\Omega\ran=\lan \Omega |\phi(x)\phi(y)|\Omega\ran\neq 0.
\ee 
Thus $S$ clearly cannot have a representation as an operator just on $A$.  Fortunately there is no reason to expect this operator to have an AdS-Rindler reconstruction, but the broader lesson is that we should really expect AdS-Rindler reconstruction to in general produce only a subalgebra of the operators on $\HC$.  We saw in section \ref{opalgsec} that the condition a subalgebra must obey for this to be possible is that the subalgebra must commute with the projection onto $\HC$ of any operator on $\ol{A}$.  In fact this is precisely the condition that we expect to be true for \textit{local} operators in $\W_C[A]$ (and their sums and products), which by bulk causality should commute with operators in $\W_C[\ol{A}]$.\footnote{This statement is rendered somewhat more subtle by the need for gravitational dressing to localize operators in a gravitational theory; we will see in section \ref{backreactsec} that this commutator continues to vanish at higher orders in $1/N$, as required by operator algebra quantum error correction.}  That this commutator vanishes with the projections onto $\HC$ of \textit{all} CFT operators in $\ol{A}$ is not something we can prove directly, but AdS-Rindler reconstruction requires it.

A second reason to prefer operator algebra quantum error correction is that even when the right hand side of \eqref{bulkcorr} is small, it will at most be suppressed by some fixed power of $1/N$.  This is because we should not smear the operators over distances shorter than the Planck length.  Since we in principle would like a version of AdS-Rindler reconstruction that works to all orders in $1/N$, it would be unsatisfying if our error correction interpretation failed at some finite order because of bulk correlation.  

We can now state our final proposal: the AdS-Rindler reconstruction of local bulk operators in \cite{Hamilton:2006az,Morrison:2014jha} is dual in the CFT to the operator algebra quantum error correction of \cite{beny2007generalization,beny2007quantum}.  An erasure of a region $\ol{A}$ is correctable if the $\phi_i(x)$'s used in defining the code subspace all lie within the causal wedge $\W_C[A]$.  In cases where the operators we are interested in are well-localized away from the causal surface $\chi_A$ of $\W_C[A]$, the situation is well-approximated by conventional quantum error correction.  Either way, the further the $\phi_i(x)$'s are from the asymptotic boundary, the better they are protected from CFT erasures.

It is worth emphasizing that in the case where a bulk operator is of order an AdS radius distance from $\W_C[\ol{A}]$, our approximate equivalence between conventional and operator algebra quantum error correction requires sub-AdS scale bulk locality.  This is a special property of those CFTs that have local holographic duals, which we have here reformulated in the language of quantum information theory.

\subsection{Disconnected regions and quantum secret sharing}\label{discsec}
\begin{figure}
\begin{center}
\includegraphics[height=4.5cm]{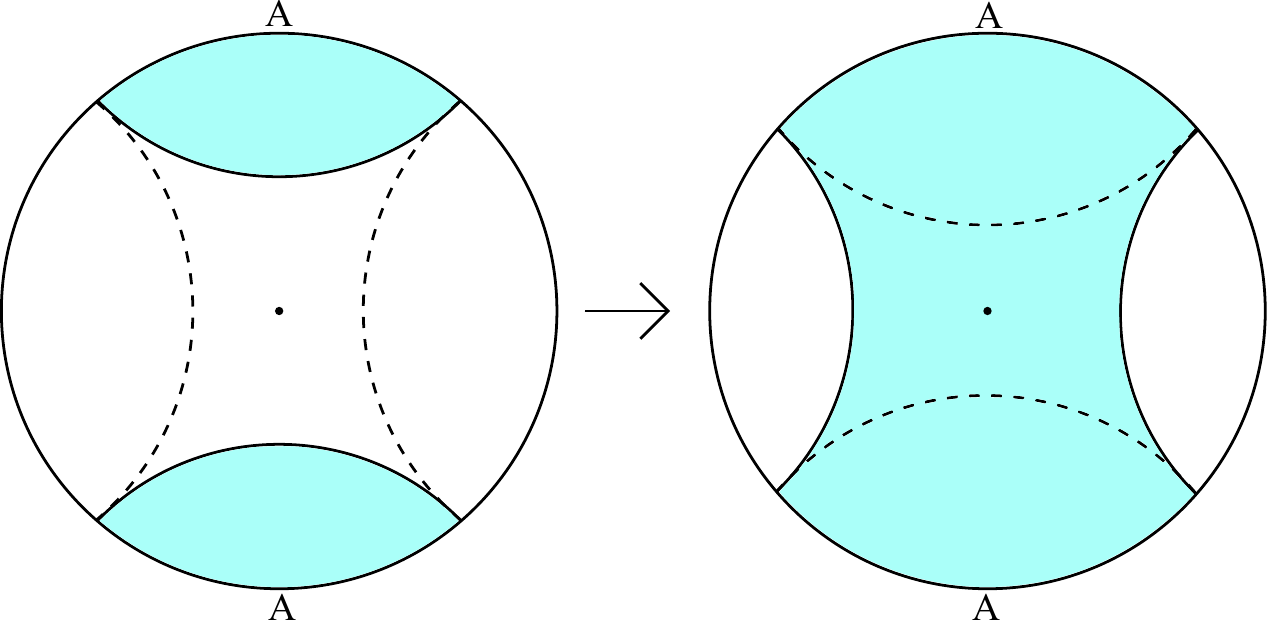}
\caption{A reconstruction phase transition?  As we increase the region $A$, the extremal-area codimension two surface of smallest area whose boundary is $\partial A$, shown as the solid lines, changes discontinuously.  Does this mean that we can now reconstruct the point in the center as an operator on $A$?}\label{multiplefig}
\end{center}
\end{figure}

So far we have only discussed the erasure of connected regions of the boundary.  More general erasures are also interesting. Consider for example  the $AdS_3$ situation depicted in figure \ref{multiplefig}. 
Here we consider a region $A$ which is the union of two disjoint intervals; in other words we have erased two disjoint intervals.  Can we choose a code subspace where we can realize the bulk operator in the center as an operator acting on $A$ or $\ol{A}$?  If the AdS-Rindler reconstruction is the last word on bulk reconstruction \cite{Bousso:2012sj}, then the answer is clearly no; this point lies neither in $\W_C[A]$ nor in $\W_C[\ol{A}]$.  This is possible within the context of quantum error correction, but only if both $A$ and $\ol{A}$ can access \textit{partial} information about the code subspace.  For example, say that $\ol{A}$ had no information whatsoever about which state of the code subspace we are in.  Then by definition \eqref{erasecond} would hold, so we could recover the information from $A$.  We are not, however, able to determine whether or not such partial information is really present.

In fact there have been recent conjectures in the literature that this operator can still be reconstructed in $A$ as long as $A$ is bigger than $\ol{A}$; more generally, the claim is that one can do reconstruction throughout the \textit{entanglement wedge}, which is defined as the bulk domain of dependence of any bulk spacelike surface whose boundary is the union of $A$ and the codimension two extremal-area surface of minimal area whose boundary is $\partial A$ \cite{Czech:2012bh,Wall:2012uf,Headrick:2014cta,Jafferis:2014lza}.\footnote{One argument that this must be the case is as follows: the extension by \cite{Faulkner:2013ana} of the Ryu-Takayanagi proposal \cite{Ryu:2006bv} to next to leading order in $1/N$ claims that bulk entanglement entropy in the entanglement wedge contributes to the Von Neumann entropy $S_A$.  So for example if we have a spin sitting in the entanglement wedge of $A$ that is entangled with another spin in $\W_C[\ol{A}]$, then the spin in the entanglement wedge contributes $\log 2$ to $S_A$.  It is hard to see how this could be the case if we could not project the spin onto a definite state by doing some measurement on $A$, but the operator we measure would then be a representation on $A$ of a logical operator on the spin.}  In the figure, the intersection of the entanglement wedge with a bulk Cauchy surface is shaded blue; the minimal area condition causes a discontinuous change as we increase the size of $A$.  Is this conjecture compatible with our proposal?  Indeed it is; we saw below equation \eqref{lcond} that in a generic code subspace \textit{any} $A$ which is greater than half of the system can correct for the erasure of its complement $\ol{A}$.\footnote{More carefully it has to be greater than half by an amount that depends on the size of the code subspace, which we can think of as being parametrically smaller than the full Hilbert space.  We explore this in more detail in the following section.}  The sharp jump in correctability as $A$ surpasses $\ol{A}$ in size is consistent with our analysis around \eqref{lcond}, where from Page's theorem we expect that the density matrix of $\ol{A}$ together with the reference system will approach being maximally mixed exponentially fast once we cross the transition.

In section \ref{qsssec} we saw that a division of the CFT into a union of shares with the property that any collection of the shares has either complete information or no information about the encoded state is called a quantum secret sharing scheme; we now see that in the situation of figure \ref{multiplefig} we will be able to reconstruct the operator in the center if and only if our boundary division into four regions gives a quantum secret sharing scheme.  

\subsection{MERA as an error correcting code?}
One shortcoming of our work so far is that, although we have laid out a plausible CFT interpretation of AdS-Rindler reconstruction as quantum error correction, we have ultimately relied on the bulk in deriving this reconstruction.  This boils down to the assumption that there exist operators in the CFT that obey the \textit{bulk} equations of motion and algebra on a subspace.  We then use this assumption to perform the Bogoliubov transformation that relates the global and the Rindler reconstructions.  This assumption is quite plausible, and essentially follows from the assumed large-$N$ structure of the CFT \cite{Heemskerk:2009pn}, but it would still be nice if we could \textit{explicitly} demonstrate the structure of the quantum error correcting code in the CFT.  In particular, in section \ref{bcs} we had to use bulk causality to argue that the necessary and sufficient condition \eqref{commreq} for operator algebra quantum error correction held, and we were not able to check it explicitly for all possible CFT operators on $\ol{A}$.  Similarly we were unable to determine whether or not the central point could be reconstructed in the two-interval $A$ of the previous subsection.  

A promising starting point for addressing these issues is the MERA tensor network construction of a discrete version of AdS/CFT \cite{Vidal:2008zz,Swingle:2009bg,evenbly2011tensor}.  It seems possible that in that fairly controlled setting one could rigorously confirm the quantum error correction structure we have motivated in this paper.  Moreover, one could attempt to determine explicitly whether or not the example of the previous subsection allows reconstruction of the operator in the center; this could be done by using the global construction to make a code subspace, entangling this code subspace with a reference system $R$ to prepare a state $|\phi\ran$, and then seeing whether there is mutual information between $R$ and $\ol{A}$. The state $|\phi\ran$ would still be prepared by a tensor network, with tensors acting both on the CFT and/or the reference system.  This calculation would go a long way towards settling the ``causal wedge vs. entanglement wedge'' debate of bulk reconstruction.\footnote{One challenge in doing this is that the region of discrepancy between the entanglement wedge and the causal wedge in that example has a size which is only of order the AdS radius, and it is not so clear how to represent sub-AdS scale physics using MERA.  One strategy for getting around this is to consider the limit of $A$ being a union of a large number of smaller disjoint intervals of equal size; this allows a parametrically large separation between the causal wedge and the entanglement wedge.}
We will not attempt this calculation here, but the typicality argument leading to \eqref{lcond} favors the entanglement wedge; we will say more about this in section \ref{countstate}.

\section{Backreaction and holography}\label{backreactsec}
We now turn to the question of how large we can make the code subspace $\HC$.  Each $\phi_i(x)$ that we act with raises the energy of the state, so doing so repeatedly will eventually lead to backreaction becoming important.  When this happens it is clear that the approximation of perturbation theory around a fixed background geometry will break down.  In this section we argue that this is related to a basic property of error correcting codes: the larger the code subspace, the fewer correctable errors.  For erasures we quantified this in equation \eqref{lcond} above.  

\subsection{Defining local operators}
\begin{figure}
\begin{center}
\includegraphics[height=5.5cm]{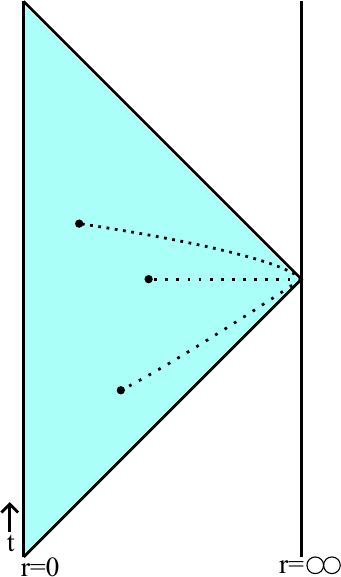}
\caption{Locating bulk points using spatial geodesics.  By construction we can only define points that lie in the bulk domain of dependence of any bulk Cauchy surface with boundary $\Sigma$.}\label{findpoints}
\end{center}
\end{figure}

Once we allow nontrivial backreaction, it is no longer possible to ignore the issue of how we define bulk local operators in a diffeomorphism-invariant way.  Following \cite{Heemskerk:2012np,Kabat:2013wga}, we do this by choosing a cutoff surface at large but finite radius, with induced metric $\mathbb{S}^{d-1}\times \mathbb{R}$, and then specifying bulk points by sending in spacelike geodesics from the $t=0$ slice of this cutoff surface that start out orthogonal to the $\mathbb{S}^{d-1}$ directions.  We then take the limit as the cutoff surface approaches the boundary.  Points are labeled by a location on $\mathbb{S}^{d-1}$, a renormalized proper distance along the geodesic, and an angle in the radial/temporal plane.  This is illustrated in figure \ref{findpoints}. 
These geodesics can be thought of as the ``gravitational dressing'' of the bulk operator, analogous to the Wilson line one would use to connect a charged operator to the boundary to make it gauge-invariant in electrodynamics.  

\begin{figure}
\begin{center}
\includegraphics[height=4cm]{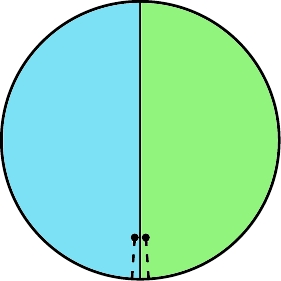}
\caption{Two dressed bulk local operators.  If the operator on the right is to be reconstructed on the boundary of the green wedge, and the operator on the left to be reconstructed on the boundary of the blue wedge, then they must be commuting operators.  By taking one of these operators to the boundary, we conclude that any dressed bulk local operator must commute with all boundary local operators \textit{except} possibly those at the endpoint of its geodesic.}\label{commutatorfig}
\end{center}
\end{figure}

As in the electromagnetic case, the operators defined in this way will have nonlocal commutators due to their nontrivial Dirac brackets.  The study of these commutators was initiated in \cite{Heemskerk:2012mn}, and more recently elaborated in \cite{Kabat:2013wga}.\footnote{A subtlety here is that they send their geodesics from arbitrary boundary times, but take them to be orthogonal to the boundary in the temporal direction as well as the $\mathbb{S}^{d-1}$ directions. These operators agree with ours at $t=0$, which is where we will study them, but as a matter of principle we have not made their choice because we want to restrict to Schrodinger-picture operators acting on a fixed time slice at the boundary.  As explained in more detail in section 4.2 of \cite{Heemskerk:2012mn}, this is the natural way to define the fixed-time Hilbert space in the bulk, since it makes it clear that expectation values are independent of possible sources on the boundary at later times.}  A full analysis has not yet been completed, however, and one point that has not yet been addressed is essential for the consistency of the AdS/Rindler reconstruction at higher orders in $1/N$: to all orders in $1/N$ perturbation theory around a fixed background, two dressed bulk operators with the property that all points on their dressing geodesics are mutually spacelike separated in that background must commute.\footnote{More carefully, we should hold the boundary endpoints of the geodesics fixed as we take $N\to\infty$ for this statement to hold.  One of us (DH) has checked this statement directly in bulk canonical gravity in the setup of \cite{Kabat:2013wga}, by computing the Dirac brackets and seeing that they are local in the $\hat{x}$ directions to any finite order in perturbation theory, but we postpone discussion of this to a future publication.}  The reason this must be the case is illustrated in figure \ref{commutatorfig}.

\begin{figure}
\begin{center}
\includegraphics[height=4cm]{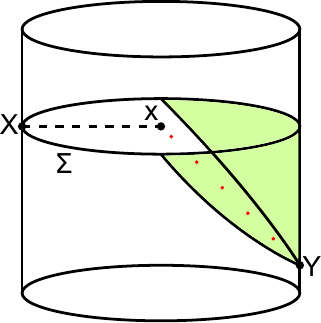}
\caption{An algebraic paradox: a bulk operator at $\phi(x)$ is dressed by a geodesic ending at $X$.  The bulk algebra suggests this operator commutes with all local CFT operators on the Cauchy surface $\Sigma$ except those at $X$, and is thus a local operator there. However, we can find a boundary point $Y$ which is both spacelike separated from $X$ and causally separated from $x$; the bulk would then require local operators there which don't commute with $\phi(x)$, while the CFT would require them all to commute.}\label{causality}
\end{center}
\end{figure}
We can use this observation to verify that bulk non-locality from the gravitational dressing of operators does not invalidate some of our previous claims.  In the introduction we argued that, because in the bulk theory a local operator in the center of the space commutes with all local operators at the boundary, the bulk operator algebra is inconsistent with the CFT algebra.  We can now give a version of this argument that includes the gravitational dressing; from figure \ref{commutatorfig}, we see that we should modify the previous statement to ``commutes with all local operators at the boundary except at one point''.  Were this to hold as an operator equation in the CFT, it would now not imply that the operator in the center must be trivial in the CFT, but it would imply that this operator can be nontrivial at $t=0$ \textit{only} at the point where the dressing geodesic ends.\footnote{As before this statement is straightforwardly true for scalars, fermions, and/or abelian gauge fields on a lattice, and we expect it to hold also for non-abelian lattice gauge fields.  In the continuum it follows from a combination of the time-slice axiom and ``Haag duality'', which says that if we split a time-slice into two regions the commutant of the set of local operators in the domain of dependence of one of the regions is the set of local operators in the domain of dependence of the other region.  In the case of gauge theories one has to be careful since Haag duality does not quite hold, since there can be a nontrivial center of the algebra for a region, but the center is always localized near the boundary \cite{Casini:2013rba} and shouldn't disrupt this argument.}  This statement, however, is not consistent with bulk causality, as we illustrate in figure \ref{causality}.  So we thus indeed find that the bulk operator algebra cannot be realized in the CFT at the level of operator equations.  As already explained, the resolution is that the bulk algebra holds in the CFT only acting on a code subspace of states.

Similarly we can now revisit our claim that bulk operators in $\W_C[A]$ perturbatively commute with bulk operators in $\W_C[\ol{A}]$, which was a necessary condition for our interpretation of the AdS-Rindler construction as operator algebra quantum error correction.  But this is exactly what the argument of figure \ref{commutatorfig} accomplishes; as long as the gravitational dressing of an operator at $x\in \W_C[A]$ also lies entirely in $\W_C[A]$, meaning that the spatial geodesic connecting $x$ to the boundary also lies in $\W_C[A]$, then it will only have non-local commutators with operators that are also located in $\W_C[A]$; any operator whose localizing geodesic is entirely in one wedge will still perturbatively commute with any operator whose localizing geodesic is entirely in the complementary wedge.

In this subsection, to connect to the formalism of \cite{Kabat:2013wga} we studied only operators attached to geodesics that start out orthogonal to the boundary time direction at $t=0$.  It would be interesting to do the analogue of their analysis at arbitrary temporal-radial angle; this amounts to working with boundary conditions that approach the ``open-FRW'' slicing of AdS
\be
ds^2=-dt^2+\sin^2 t \left(d\chi^2+\sinh^2 \chi d\Omega_{d-1}^2\right)  
\ee
as $\chi\to \infty$.  As explained in \cite{Heemskerk:2012mn}, this would be a natural bulk construction of Schrodinger picture gauge-invariant operators on the fixed-boundary-time Hilbert space.

\subsection{Shrinking of the causal wedge}
We now return to the question of how backreaction affects causal wedge reconstruction.  Our basic proposal is that adding energy in the bulk causes the causal wedge of a fixed boundary region $A$ to recede towards the boundary, giving it less access to bulk operators defined at fixed renormalized geodesic distance (for some related discussion see \cite{Wall:2012uf,Hubeny:2012wa,Hubeny:2012ry,Blanco:2013joa,Headrick:2014cta}).  

Consider for example the AdS-Schwarzschild geometry in $d+1$ dimensions.
\be
ds^2=-f(r)dt^2+\frac{dr^2}{f(r)}+r^2 d\Omega_{d-1}^2,
\ee
with 
\be
f(r)\equiv r^2+1-\frac{\alpha}{r^{d-2}}.
\ee
$\alpha$ is proportional to the ADM mass of this geometry.  Now consider a boundary disc $A$ of angular size $\theta$; its causal wedge reaches a radius $r_\theta(\alpha)$ in the bulk defined implicitly by
\be\label{rdef}
\frac{\theta}{2}=\int_{r_\theta(\alpha)}^\infty \frac{dr'}{f(r')}.
\ee
The proper distance of this radius to a cutoff surface at $r=r_c$ is 
\be
\int_{r_\theta(\alpha)}^{r_c}\frac{dr'}{\sqrt{f(r')}}=\int_{r_\theta(\alpha)}^{r_c}dr'\left(\frac{1}{\sqrt{f(r')}}-\frac{1}{r'}\right)+\log\frac{r_c}{r_\theta(\alpha)},
\ee
so we can subtract $\log r_c$  to define a renormalized proper distance
\be\label{ddef}
d_\theta(\alpha)\equiv \int_{r_\theta(\alpha)}^{\infty}dr'\left(\frac{1}{\sqrt{f(r')}}-\frac{1}{r'}\right)-\log r_\theta(\alpha).
\ee
We claim that $d_\theta(\alpha)$ is a decreasing function of $\alpha$ at fixed $\theta$, which by differentiating under the integral sign is equivalent to the claim that
\be
\int_{r}^\infty \frac{dr'}{r'^{d-2}}\frac{1}{f(r')^{3/2}}\left[\sqrt{\frac{f(r)}{f(r')}}-\frac{1}{2}\right]>0
\ee
for all $\alpha>0$ and for all $r>r_+(\alpha)$, where $r_+(\alpha)$ is the positive root of $f$.  This can be shown analytically in various limits, and is easily checked numerically in the general case.  One can also study the asymptotically-$AdS_3$ BTZ black hole, where a similar result holds and all integrals can be done analytically.  Thus we see that indeed the causal wedge has access to fewer and fewer bulk observables as we increase the mass of the matter in the center.  This after all must be the case, since as we keep increasing the mass a point at fixed renormalized geodesic distance from the boundary will eventually go through the horizon.

It is interesting to think about how general this statement is; under what circumstances can the causal wedge move \textit{inwards} in renormalized geodesic distance as we insert energy?  One might guess that the null energy condition should generically prevent this, but to test that we need a more precise conjecture.  One first guess is that in any geometry obeying the null energy condition the causal wedge of a fixed boundary region can see at most as far in renormalized geodesic distance as it can in the vacuum.  In fact this conjecture is false, we have constructed explicit counterexamples.  Indeed a weaker conjecture, where we replace the null energy condition by the dominant energy condition\footnote{With the presence of a negative cosmological constant, we only impose the dominant energy condition on the ``matter'' part of stress tensor which does not include the cosmological constant.  This was used in \cite{Gibbons:1982jg} to prove the positive mass theorem in asymptotically anti-de Sitter space.}, still has counterexamples.  One counterexample is given by a small perturbation of $AdS_4$, with the metric
\be
ds^2 = -f(r) dt^2 + \frac{dr^2}{f(r)} + r^2 d\Omega_2^2 \,,
\ee
where
\be
f(r) = r^2 + 1 + \epsilon h(r) \,,\qquad
h(r) = -\frac{r^2}{\left(\sqrt{r^2+1}+10\right)^3} \,.
\ee
With a small positive $\epsilon$, the causal wedges of certain fixed boundary regions can see farther in renormalized geodesic distance than they can in the vacuum.  These boundary regions include spherical regions whose causal wedges probe deep into the bulk geometry.

Although these counterexamples prevent any straightforward ``monoticity of causal wedge recession theorem'', we expect that the Schwarzschild calculation we have just discussed captures the general tendency.  It would be nice to prove a more general theorem verifying this, but we have not succeeded in finding one.

\subsection{Counting states}\label{countstate}
The recession of the causal wedge has a nice quantum error correction interpretation; as we allow the code subspace to have more and more excited states, a bulk operator localized at some fixed geodesic distance will eventually no longer lie in the causal wedge of a fixed boundary region.  In other words, the code will lose some of its ability to correct erasures; we will need access to more of the boundary to study the same bulk observables.   In this subsection we study this a bit more quantitatively, making contact with the general condition \eqref{lcond} for typical correctability.  

To apply \eqref{lcond} to AdS/CFT, we need to identify CFT analogues of the quantities $n$, $l$, and $k$.  $n$ is the total number of qubits used in doing the encoding, and should roughly correspond to the total number of CFT degrees of freedom relevant for reconstructing a particular bulk region of interest.  This is somewhat nontrivial; the CFT has an infinite number of degrees of freedom in the UV which are needed to reconstruct bulk operators that are arbitrarily close to the boundary.  To deal with this we take our code subspace to only involve states where we act on the vacuum with operators $\phi_i(x)$ that are all localized within a region $\mathcal{R}$ at the center of the AdS space that has proper size of order the AdS radius.  We will also take them to be smeared over distances that are small compared to their separation from the boundary of $\mathcal{R}$, so that we do not have to worry about the difference between conventional and operator algebra quantum error correction.  The global reconstructions \eqref{smear} of these operators involve integrals over functions that vary smoothly on the scale of the radius of curvature of the boundary $\mathbb{S}^{d-1}$, so we can integrate out all CFT degrees of freedom with shorter wavelength.\footnote{It is not hard to generalize to the case where $\mathcal{R}$ is taken to be parametrically larger than the AdS radius, this essentially involves a repeat of the analysis of \cite{Susskind:1998dq}.}  For concreteness we will consider the case of the $\mathcal{N}=4$ super Yang-Mills theory in $3+1$ boundary dimensions with gauge group $SU(N)$, in which case we have
\be
n\sim N^2.
\ee
Erasing a disc of angular size $\theta$ will then correspond to erasing 
\be
l=\frac{\theta- \sin \theta}{2\pi} n
\ee
qubits, where this function is just $n$ times the ratio of the area of the disc to the area of the $\mathbb{S}^3$.    

Let us first consider the case where the code subspace is small, that is when $k\sim 1$.  From \eqref{lcond} we then expect that we can correct for the erasure as long as $n-2l \gg 1$, or in other words $\theta<\pi$.  But this is exactly what we expect from the AdS-Rindler reconstruction; once $\theta<\pi$, $W_C[A]$ will contain the center of the space.  It is interesting to note that the derivation of \eqref{lcond} applied to an erasure of an \textit{arbitrary} collection of $\ell$ qubits, so this suggests that we should also be able to reconstruct operators in the center on a union of disconnected regions, provided that together they make up more than half of the boundary.  With regards to our discussion of section \ref{discsec}, this gives support to the entanglement wedge over the causal wedge.

We can now start increasing $k$; nothing interesting will happen until we get $k\sim N^2$, after which the set of erasures we are able to correct will start decreasing.  But this is exactly the condition for backreaction to become important in the center; with $k\sim N^2$ the entropy of the code subspace is comparable to that of a black hole filling $\mathcal{R}$ and thus most states in the code subspace must actually be black holes.  So both on the CFT side through equation \eqref{lcond} and the bulk side via backreaction we arrive at the same conclusion for when correctability should break down.  This is a manifestation of the holographic entropy bound of \cite{Susskind:1994vu}.  

\section{Conclusion}
In this paper we have provided what we consider to be a new understanding of how the holographic principle is realized in AdS/CFT.  Bulk effective field theory operators emerge as a set of logical operations on various encoded subspaces, which are protected against local errors in the boundary CFT.  The bulk algebra is realized only on these subspaces, and only if we do not try to describe too many operations at once.  Asking for more causes the error correction procedure to fail, which in the bulk is manifested by the formation of a black hole.  

To some extent we have only recast known facts about the AdS-Rindler reconstruction in a new language, but in our view that construction is quite opaque once the operators in the boundary domain of dependence of $A$ are evolved back to the boundary Cauchy surface $\Sigma$ at $t=0$.  Our description in terms of error correction is phrased entirely on this Cauchy surface, and gives what we feel to be a satisfying interpretation of how the AdS-Rindler reconstruction is realized in the CFT that cleanly resolves some of its paradoxical features.    

It is of course interesting to ask if there are any implications of this work for the recent controversy on whether or not the interiors of black holes are describable in AdS/CFT; for now we leave this for future study.

\paragraph{Acknowledgments} We'd like to thank Vijay Balasubramanian, Will Donnelly, Ethan Dyer, Patrick Hayden, Jennifer Lin, Juan Maldacena, Don Marolf, Eric Mintun, Ian Morrison, Joe Polchinski, Vladimir Rosenhaus, Steve Shenker, Douglas Stanford, Lenny Susskind, Herman Verlinde, Bob Wald, and Aron Wall for useful discussions.  XD is supported by the National Science Foundation under grant PHY-0756174, and DH is supported by the Princeton Center for Theoretical Science.  We are also grateful to the Aspen Center for Physics for providing a stimulating research environment during the ``Emergent Spacetime in String Theory'' workshop, where this work was initiated.  The center is funded in part by the NSF grant no.\ PHYS-1066293.

\appendix
\section{More details on AdS-Rindler reconstruction}\label{rindapp}
In this appendix we review a bit more about the AdS-Rindler reconstruction.  The AdS-Rindler wedge has metric
\be
ds^2=-(\rho^2-1)d\tau^2+\frac{d\rho^2}{\rho^2-1}+\rho^2 d H_{d-1}^2,
\ee 
where $d H_{d-1}^2$ is the standard metric on the $d-1$ dimensional hyperbolic ball $H_{d-1}$.  We will refer to coordinates on $H_{d-1}$ collectively as $\alpha$.  A free real scalar field on this background can be expressed in the Heisenberg picture as
\be\label{phieq}
\phi(\rho,\tau,\alpha)=\int_0^\infty\frac{d\omega}{2\pi}\sum_\lambda\left(f_{\omega \lambda}(\rho,\tau,\alpha) a_{\omega\lambda}+f_{\omega \lambda}^*(\rho,\tau,\alpha) a_{\omega\lambda}^\dagger\right),
\ee
where $f_{\omega \lambda}(\rho,\tau,\alpha)$ is a solution of the Klein-Gordon equation of the form
\be
f_{\omega \lambda}(\rho,\tau,\alpha)=e^{-i\omega \tau} Y_\lambda(\alpha) \psi_{\omega\lambda}(\rho).
\ee
Here $Y_\lambda(\alpha)$ is an eigenfunction of the Laplacian on $H_{d-1}$ with eigenvalue $-\lambda$; the set of $\lambda$'s is continuous (and positive) so ``$\sum_\lambda$'' should really be understood as shorthand for an integral together with a sum over degeneracies.  $\psi_{\omega \lambda}$ is explicitly given by
\begin{align}\nonumber
\psi_{\omega,\lambda}(\rho)=\mathcal{N}_{\omega\lambda}\rho^{-\Delta}\left(1-\frac{1}{\rho^2}\right)^{-\frac{i\omega}{2}}F\Bigg(&-\frac{(d-2)}{4}+\frac{\Delta}{2}-\frac{i\omega}{2}+\frac{1}{2}\sqrt{\frac{(d-2)^2}{4}-\lambda},\\
&-\frac{(d-2)}{4}+\frac{\Delta}{2}-\frac{i\omega}{2}-\frac{1}{2}\sqrt{\frac{(d-2)^2}{4}-\lambda},\Delta-\frac{d-2}{2},\frac{1}{\rho^2}\Bigg),
\end{align}
with 
\be
\mathcal{N}_{\omega \lambda}=\frac{1}{\sqrt{2|\omega|}}\frac{\Gamma\left(-\frac{(d-2)}{4}+\frac{\Delta}{2}+\frac{i\omega}{2r_s}+\frac{1}{2}\sqrt{\frac{(d-2)^2}{4}-\frac{\lambda}{r_s^2}}\right)\Gamma\left(-\frac{(d-2)}{4}+\frac{\Delta}{2}+\frac{i\omega}{2r_s}-\frac{1}{2}\sqrt{\frac{(d-2)^2}{4}-\frac{\lambda}{r_s^2}}\right)}{\Gamma\left(\Delta-\frac{d-2}{2}\right)\Gamma\left(\frac{i\omega}{r_s}\right)}
\ee
and
\be
\Delta=\frac{d}{2}+\frac{1}{2}\sqrt{d^2+4m^2}.
\ee
The normalization is chosen to ensure that $a$ and $a^\dagger$ have the usual algebra, and implies that for $\rho$ large and negative, $\psi_{\omega \lambda}$ is of order $\frac{1}{\sqrt{\omega}}$.  $F$ is a hypergeometric function that goes to one as $\rho \to \infty$.  By comparing to the extrapolate dictionary \eqref{extd} we can read off that 
\be
\mO_{\omega \lambda} = \int d\tau d\alpha e^{i \omega \tau}Y_\lambda^*(\alpha)\mO(\tau,\alpha)=N_{\omega \lambda} a_{\omega \lambda},
\ee
and substituting back into \eqref{phieq} and formally exchanging the $\tau\alpha$ integral with the $\omega \lambda$ integral/sum we arrive at
\be
\phi(\rho,\tau,\alpha)=\int d\tau' d\alpha' K(\rho,\tau,\alpha; \tau'\alpha') \mO(\tau',\alpha').
\ee
The ``smearing function'' $K$ is given by
\be\label{rindK}
K(\rho,\tau,\alpha; \tau'\alpha')=\int_{-\infty}^\infty \frac{d\omega}{2\pi}\sum_\lambda \frac{1}{N_{\omega \lambda}}f_{\omega \lambda}(\rho,\tau,\alpha)e^{i\omega \tau'}Y^*_\lambda(\alpha')e^{i\omega \tau'}.
\ee
$K$ can be understood as a kernel for constructing a bulk solution of the KG equation in the AdS-Rindler wedge given arbitrary boundary conditions at spatial infinity as a function of $\tau$ and $\alpha$.  It was immediately realized, however, that in fact this expression for $K$ is not well-defined \cite{Hamilton:2006az} (see also \cite{Leichenauer:2013kaa,Rey:2014dpa}); the reason is that from Stirling's formula it is straightforward to see that at large $\lambda$ we have
\be
|N_{\omega \lambda}|^2\sim e^{-\pi \sqrt{\lambda}}.
\ee 
This is problematic for the convergence of the $\lambda$ integral in \eqref{rindK}, and using the WKB approximation for $Y$ and $\psi$ at large $\lambda$ it is not hard to see that indeed the integral does not converge for any choice of bulk and boundary points \cite{Leichenauer:2013kaa,Rey:2014dpa,Morrison:2014jha}.

\begin{figure}
\begin{center}
\includegraphics[height=4cm]{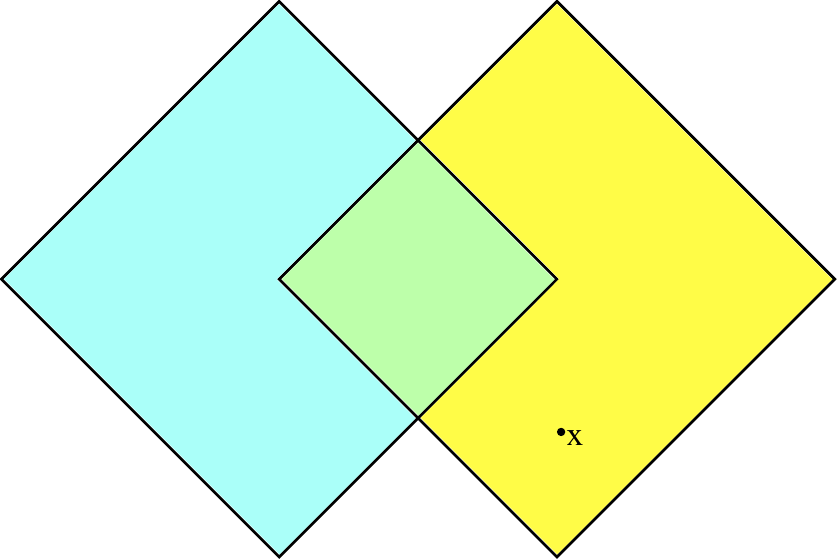}
\caption{Overlapping Rindler wedges, shown in the boundary.  $D[A]$ is in blue, $D[B]$ is in yellow, and their overlap is in green.}\label{overlap}
\end{center}
\end{figure}
In fact, the nonconvergence of $K$ is necessary to avoid the following paradox; say that we have two overlapping Rindler wedges, $W_C[A]$ and $W_C[B]$, as in figure \ref{overlap}.  If $K$ existed then since the $f_{\omega\lambda}$ solutions are complete it would construct a unique solution of the KG equation in $W_C[A]$ with some particular boundary conditions in $D[A]$.  We can, however, imagine modifying the spatial boundary conditions at a point $x$ that is in $D[B]$ but not in $D[A]$, such that $x$ is nonetheless causally separated from a point in $W_C[A]$.  We then should be able to send a signal to $W_C[A]$ without modifying the boundary conditions in $D[A]$, which contradicts the uniqueness of the solution.  

In \cite{Hamilton:2006az} it was argued that one should analytically continue in the boundary spatial coordinates to avoid this divergence, but this is a rather unusual thing to do to a quantum field theory and would be rather problematic from the point of view of our analysis in this paper.  The issue was recently illuminated considerably in \cite{Morrison:2014jha} (see also \cite{Papadodimas:2012aq}), where it was argued that as long as we are careful to think of $K$ as a distribution for integration against CFT correlation functions we are allowed to use it without any analytic continuations.  The key point of \cite{Morrison:2014jha} was that at least at leading order in $1/N$, if we integrate $K$ against appropriate bulk test functions then its singularity structure is such that we are always able to integrate it against CFT expectation values
and get a reasonable answer.  Intuitively, the reason that this is able to avoid the contradiction of the previous paragraph is that in the CFT $\mO$ obeys a boundary equation of motion; we are not free to choose it independently at different boundary times.  Turning on a source at $x$ will necessarily propagate in the boundary into $D[A]\cup D[B]$, so we will not be able to change the boundary data at $x$ without also changing it in $D[A]$.    

The argument of \cite{Morrison:2014jha} does not immediately generalize to higher order corrections in $1/N$, but we expect a more detailed analysis will show that it can be improved order by order in $1/N$.\footnote{Recently it was argued that AdS/CFT in the AdS-Rindler wedge cannot be understood as a statement about subregions in the global CFT \cite{Chowdhury:2014oba}.  It is true that there is a subtlety here in that the cutoffs in the bulk are different in the two cases, but AdS/CFT is really a statement about the continuum CFTs.  The discrepancies discussed in \cite{Chowdhury:2014oba} were localized to ``cutoff sized regions'' in the CFT, which really means that they are not part of the continuum theory.  That CFT expectation values have the right singularity structure to be integrated against $K$ in \eqref{rindK} is related to the fact that they are computed in states that remain finite energy as we take the continuum limit, and as explained in \cite{Morrison:2014jha} it is forgetting this that leads to trouble.}
 
\section{The basic theorem of operator algebra quantum error correction} \label{thmapp}
In this appendix we prove the theorem of section \ref{opalgsec}.  The proof is original, but the theorem is a special case of the results of \cite{beny2007generalization,beny2007quantum}.
\begin{proof}
Indeed say that we have a code subspace $\HC$ spanned by an orthonormal basis $|\wt{i}\ran_{E\ol{E}}$.  Moreover say that $O$ is an operator which acts as
\begin{align}\nonumber
O|\wt{i}\ran&=\sum_j O_{ji}|\wt{j}\ran\\
O^\dagger|\wt{i}\ran&=\sum_j O_{ij}^*|\wt{j}\ran.
\end{align}
Finally, say that for any operator $X_E$ acting on $E$ we have
\be\label{commcond}
\lan \wt{i}|[O,X_E]|\wt{j}\ran=0  \qquad \qquad \forall  i,j.
\ee
We'd then like to show that there exists an operator $O_{\ol{E}}$ acting only on $\ol{E}$ and obeying
\begin{align}\nonumber
O_{\ol{E}}|\wt{i}\ran&=O|\wt{i}\ran\\
O_{\ol{E}}^\dagger|\wt{i}\ran&=O^\dagger |\wt{i}\ran.\label{goal}
\end{align}

Let us first observe that in general in a bipartite system in a state
\be
|\psi\ran=\sum_{ab}C_{ba}|a\ran_A |b\ran_B,
\ee
operators $O_A$ on $A$ and $O_B$ on $B$ will obey 
\be
O_B|\psi\ran=O_A|\psi\ran
\ee
if and only if 
\be\label{inverse}
O_B C=C O_A^T.
\ee


Now in the setup of the theorem, let us consider the state\footnote{For convenience we have dropped the overall normalization, it will cancel between the two sides of \eqref{inverse}.}
\be\label{phidef}
|\phi\ran=\sum_i|i\ran_R |\wt{i}\ran_{E\ol{E}}.
\ee
The properties \eqref{goal} will clearly hold if and only if they hold on $|\phi\ran$, ie if
\begin{align}\nonumber
O_{\ol{E}}|\phi\ran&=O|\phi\ran\\
O_{\ol{E}}^\dagger|\phi\ran&=O^\dagger |\phi\ran,\label{goal2}
\end{align}
so this is all we need to show (\eqref{goal} follows from applying projections to $R$).  

Our strategy then is to notice that we can decompose $RE\ol{E}$ into a bipartite system in two different ways, either as the $R$ system and the $ E \ol{E}$ system or as the $\ol{E}$ system and the $RE$ system.  The first decomposition is manifest in \eqref{phidef}, and since in this case $C$ is just the identity we see that \eqref{inverse} holds and we have
\be
O_R|\phi\ran=O|\phi\ran,
\ee
where $O_R\equiv O^T$ acts just on $R$.  The point then is that we can interpret $O_R$ as $O_R\otimes I_E$, and then see if we can mirror it back onto $\ol{E}$ using \eqref{inverse}.  If so then we succeed in constructing $O_{\ol{E}}$.

To proceed, we can define
\be
|\wt{i}\ran=\sum_{a,m}C^{i}_{am}|m\ran_E |a\ran_{\ol{E}},  
\ee
in which case we have
\be
|\phi\ran=\sum_{i,a,m}C^i_{am}|im\ran_{RE}|a\ran_{\ol{E}}.  
\ee
It will be very convenient in what follows to treat $C$ as a rectangular matrix, with $a$ being the first index and $im$ being the second index.  For example, we have the reduced density matrices
\begin{align}\nonumber
\rho_{\ol{E}}&=CC^\dagger\equiv g\\
\rho_{RE}&=C^T C^*.
\end{align}
Since $g$ is a non-negative hermitian matrix with positive trace, it will be invertible on a subspace of $\ol{E}$.  Moreover, any state orthogonal to this subspace will also be orthogonal to all of the $|\wt{i}\ran$'s, so we can just take $\ol{E}$ to be given by this subspace; $g$ is then invertible.  This then means that $C$ has a right inverse $ C^\dagger g^{-1}$.

Now observe that the commutator condition \eqref{commcond} implies that
\be\label{Ccomm}
C^\dagger C O=O C^\dagger C,
\ee
or equivalently that
\be
[O_R\otimes I_E,\rho_{RE}]=0.
\ee
If $C^\dagger g^{-1}$ were a \textit{left} inverse of $C$, then \eqref{inverse} would hold if we define
\be
O_{\ol{E}}=CO C^\dagger g^{-1},
\ee
which would then give us the first equation in \eqref{goal2} (remember that $O_R=O^T$).  In fact it is a right inverse, but we can use \eqref{Ccomm} to show that \eqref{inverse} holds nonetheless:\footnote{Intuitively this is because if we do a Schmidt decomposition of $|\phi\ran$ into $RE$ and $\ol{E}$, we can choose a basis where $C$ is actually a square matrix, so there is no difference between its left and right inverses.  The condition \eqref{Ccomm} ensures us that $O_R$ acts within the subspace defined by the Schmidt decomposition.}
\begin{align}\nonumber
O_{\ol{E}}C&=CO C^\dagger g^{-1} C\\\nonumber
&=g^{-1}CC^\dagger CO C^\dagger g^{-1} C\\\nonumber
&=g^{-1}COC^\dagger C C^\dagger g^{-1} C\\\nonumber
&=g^{-1}COC^\dagger C\\\nonumber
&=g^{-1}CC^\dagger C O\\
&=CO.
\end{align}

Moreover
\begin{align}\nonumber
O_{\ol{E}}^\dagger&=g^{-1}C O^\dagger C^\dagger\\\nonumber
&=g^{-1}C O^\dagger C^\dagger C C^\dagger g^{-1}\\
&=C O^\dagger C^\dagger g^{-1},
\end{align}
where we've used \eqref{Ccomm}, so the second equation in \eqref{goal2} is also satisfied.  This concludes the proof.
\end{proof}

\bibliographystyle{jhep}
\bibliography{bibliography}
\end{document}